\documentstyle[prl,aps]{revtex}

\def\be{\begin{equation}}
\def\ee{\end{equation}}
\def\ben{\begin{eqnarray}}
\def\een{\end{eqnarray}}
\newcommand{\bei}{\begin{itemize}}
\newcommand{\eei}{\end{itemize}}
\def\ra{\rangle}
\def\la{\langle}
\def\blacksquare{\vrule height 4pt width 3pt depth2pt}
\def\pcal{{\cal P}}
\def\hcal{{\cal H}}
\def\trace{\mbox{Tr}}

\begin{document}
\draft
\twocolumn

\title{Limits for entanglement measures}

\author{Micha\l{} Horodecki$^{1}$,
Pawe\l{} Horodecki$^{2}$ and Ryszard Horodecki
$^{1}$}

\address{$^1$ Institute of Theoretical Physics and Astrophysics,
University of Gda\'nsk, 80--952 Gda\'nsk, Poland,\\
$^2$Faculty of Applied Physics and Mathematics,
Technical University of Gda\'nsk, 80--952 Gda\'nsk, Poland
}

\maketitle

\begin{abstract}
Basic principle of entanglement processing says that {\it entanglement
cannot increase under local operations and classical communication}. Basing on
this principle we show that {\it any} entanglement measure $E$
suitable for the regime  of high number of identically prepared entangled pairs satisfies
$E_D\leq E\leq E_F$ where $E_D$ and $E_F$ are entanglement of distillation
and formation respectively. Moreover, we exhibit a theorem establishing
a very general form of bounds for distillable entanglement.
\end{abstract}


Since the pioneering papers \cite{conc,huge,Bennett_pur} on
quantifying entanglement much have been done in this field 
\cite{RP,Knight,VP98,Rains_bound1,VT99,Vidal_mon,Rains_rig,Rains_bound2,WH}.
However, in the case of mixed states, we are still at the
stage of gathering phenomenology. In the very fruitful axiomatic approach
\cite{RP,Knight,VP98} there is not even  an agreement as to what
postulates should be satisfied by candidates for entanglement measures.
Moreover, we do not know the quantum communication meaning of the
known measures apart from entanglement of formation $E_F$ and entanglement
of distillation $E_D$ \cite{huge},  having the following
dual meaning:
\bei
\item $E_D(\varrho)$ is the maximal number of singlets that can be produced
from the state $\varrho$ by means of local operations and classical
communication (LQCC).
\item $E_F(\varrho)$ is the minimal number of singlets needed to produce
the state $\varrho$ by LQCC operations. 
\eei
(More precisely: $E_D$ ($E_F$) is minimal number of singlets {\it per copy} 
in the state $\varrho$ in the asymptotic sense of considering $n\rightarrow
\infty$ copies altogether).
Unfortunately, they are very hard to deal with. One can ask
a general question: Is there a rule that would somehow order the many 
possible measures satisfying some reasonable axioms? Moreover, is there any
connection between  the axiomatically defined measures  and the entanglement
of distillation and formation?

Surprisingly, it appears that just  the two, historically first,
measures of entanglement \cite{huge} constitute the sought after rule, being
{\it extreme} measures.
In this paper we show that any measure satisfying certain natural axioms
(two of them specific to the {\it asymptotic regime} of high number
of identically prepared entangled pairs) must be confined
between $E_D$ and $E_F$:
\be
E_D\leq E\leq E_F.
\ee
The result is compatible with some earlier results in this direction.
In Ref. \cite{VP_contem} Vedral and Plenio
provided heuristic argumentation that an additive measure of entanglement
should be no less than $E_D$. Uhlmann showed that non-regularized 
entanglement of formation (closely related to $E_F$ \cite{huge})
is upper bound for all convex functions that agree with it on pure
states \cite{Uhlmann}. Finally, the presented result is compatible
with the result
by Popescu and Rohrlich \cite{RP}, completed by Vidal \cite{Vidal_mon},
stating uniqueness of the entanglement measure for pure states.

The proof of the result (contained in Theorem 1) is very simple, but
it is very powerful. Indeed, as a by-product we obtain
(Theorem 2)  surprisingly weak conditions  for a function  to be
upper bound for $E_D$. This is remarkable result, as evaluation of
$E_D$ is one  of the central tasks of the present stage of quantum
entanglement theory.
In particular, we obtain elementary proof that the
relative entropy entanglement $E_r$ \cite{Knight,VP98} and the
function considered by Rains \cite{Rains_bound2} are bounds for distillable
entanglement. Note that the  proof of Ref. \cite{Rains_bound2}
involves  complicated mathematics, while the one of Ref. \cite{VP98}
is based on
still unproven additivity assumption. In addition, our result
is very general, and we expect it will result in an easy search for bounds
on distillable entanglement. It is crucial that the basic tool we
employ to obtain the results  is the  fundamental principle of
entanglement theory stating that {\it entanglement cannot increase under local
operations and classical communication} \cite{conc,huge,RP}.
Thus the principle, putting bounds for the efficiency of distillation, 
plays a similar role to that of second law of thermodynamics (cf. \cite{RP})
the basic restriction for efficiency of heat engines.

Let us first set the list of postulates we impose for entanglement measure.
So far, the rule of choosing some postulates and discarding others was
an intuitive understanding of what entanglement is. Now, we would like to add a
new rule:
{\it Entanglement of distillation is a good measure}. Thus, we cannot accept
a postulate that is not satisfied by $E_D$. This is reasonable because
$E_D$ has a direct sense of quantum capacity
of the teleportation \cite{Bennett_tel} channel constituted by the source
producing bipartite systems.
We will  see  that  this rule will suppress some of hitherto accepted
postulates: This is the lesson given us by existence of bound entangled
states \cite{bound}.

We  split the postulates into the following three groups:

{\parindent=0pt
\vskip1mm
1. {\it Obvious postulates.-}
\bei
\item[a)] Non-negativity: $E(\varrho)\geq0$;

\item[b)] vanishing on separable states: $E(\varrho)=0$ if $\varrho$ is
separable;

\item[c)] normalization: $E(|\psi_+\ra\la \psi_+|)=1$, where
$\psi_+={1\over\sqrt2}(|00\ra+|11\ra)$.
\eei

2. {\it Fundamental postulate: monotonicity under LQCC operations.-}
\bei
\item[a)] Monotonicity  under local operation:
If either of the parties sharing the pair in the state $\varrho$ performs
the operation leading to state $\sigma_i$ with probability $p_i$, then the
expected entanglement cannot increase
\[
E(\varrho)\geq \sum_ip_iE(\sigma_i);
\]
\item[b)] convexity (monotonicity under discarding information)
\[
E\left(\sum_ip_i\varrho_i\right)\leq \sum_ip_iE(\varrho_i).
\]
\eei
\vskip-1mm
3. {\it Asymptotic regime postulates.-}
\bei
\item[a)] Partial additivity
\[
E(\varrho^{\otimes n})=nE(\varrho);
\]
\item[b)] continuity: If $\la \psi^{\otimes n}|\varrho_n|\psi^{\otimes n}\ra
\rightarrow 1$ for $n\rightarrow \infty$ then
\[
{1\over n} |E(\psi^{\otimes n})-E(\varrho_n)|\rightarrow 0,
\]
where $\varrho_n$ is some joint  state of $n$ pairs.
\eei
}
Let us now briefly discuss the considered postulates. In the first group, the
postulate of normalization is to prevent us from the many trivial
measures given by positive constant multiply of some measure  $E$. The axiom
1a) is indeed obvious (separable state contains no entanglement). What, however,
is not obvious, is: Should not we require  vanishing of $E$ if and {\it only}
if the state is separable? The latter seems reasonable, because if the state
is not separable, it contains entanglement, that should be indicated by the 
entanglement measure. However, according to our rule, we should
look at distillable entanglement. We then  can see  that the bound entangled
states \cite{bound} are
entangled, but have $E_D$ equal to zero. Thus we should accept
entanglement measures that indicate no entanglement for some entangled states.
This curiosity is due to existence of different types of entanglement.

Let us now pass to 
 the second group.
The fundamental postulate, displaying the basic feature of entanglement (that
creating entanglement requires {\it global} quantum interaction) was 
introduced
in  Ref. \cite{conc,huge} and developed in  Refs. \cite{RP,Knight,VP98}. It
was put into  the above, very convenient form in Ref. \cite{Vidal_mon}. Any
function satisfying it must be invariant under product unitary
transformations and constant on separable states  \cite{Vidal_mon}.
It also follows that if a trace preserving map $\Lambda$ can be 
realized  as a
LQCC operation, then $E(\Lambda(\varrho))\leq E(\varrho)$.

The postulates of the first and the second groups are commonly accepted. The
functions that satisfied  them (without normalization axiom)
have been called {\it entanglement monotones} \cite{Vidal_mon}. 

Let us now discuss the last group of postulates, called ``asymptotic
regime'' ones because they are necessary in the  limit of large
number of identically prepared entangled pairs, and can be discarded if a small
number of pairs are considered.  This asymptotic regime is extremely important
as  it is natural regime both for the directly related
theory of quantum channel capacity \cite{huge} and the recently developed
``thermodynamics of entanglement''  \cite{RP,VP_contem,termod}.

Partial additivity says that if we have a stationary,
memoryless source, producing pairs in the state $\varrho$, then the
entanglement content  grows linearly with the number of pairs.
A plausible argument to accept this postulate was given  in Ref. \cite{RP}
in the context of thermodynamical analogies.
Vedral and Plenio \cite{VP_contem} considered full additivity
$E(\varrho\otimes \sigma)=E(\varrho)+E(\sigma)$ as
a desired property. However, the effect of activation of bound entanglement
\cite{aktyw}  suggests that $E_D$ is not fully additive, so,
according to our rule, we will not impose this stronger additivity.

Let us now pass to the last property. It says that
in the region close to the pure states, our measure is to behave regularly:
If the joint state of large number pairs is close to the product of pure
states, then the {\it densities} of entanglement (entanglement per pair)
of both of the states should be close to each other, too. This is a very
weak form of the continuity exhibited e.g. by von Neumann  entropy
that follows from  Fannes inequality
\cite{Fannes}. We do not require the latter, strong continuity, because
we expect that entanglement of distillation can exhibit some peculiarities
at the boundary of the set of bound entangled states. However, it can be seen
that $E_D$ satisfies this weak continuity displayed as the last postulate
of our list.

The continuity property as a potential
postulate for entanglement measures was considered by Vidal \cite{Vidal_mon}
in the context of the problem of uniqueness of entanglement measure for pure
states. Namely, Popescu and Rohrlich \cite{RP} starting from
thermodynamical analogies,
argued that entanglement of formation (equal to entanglement of
distillation for
pure states \cite{conc}) is a unique measure, if one imposes additivity and
monotonicity (and, of course, normalization). Later on, many monotones
different than $E_F$ on pure states
were designed \cite{Knight,VP98,VT99}. There was still no contradiction
because  they were not additive. However, Vidal constructed a set of
monotones additive for pure states, that still differed from $E_F$ for
pure states \cite{Vidal_mon}.
He removed the contradiction by pointing out the missing assumption being
just the considered continuity. The completed-in-this-way
{\it uniqueness theorem} states that a function satisfying the listed axioms
must be equal to entanglement of formation on the pure states.

In the following we will show that the above theorem can be viewed as a
special case of the general property of entanglement measures (in this paper
the functions satisfying the list of postulates we will call entanglement
measures). Before we state the theorem we need definitions of entanglement
of distillation and formation. We accept the following definitions.

$E_F$ is a regularized version of the original entanglement of formation
$E_f$ \cite{huge} defined as follows. For pure states $E_f$ is equal to
entropy of entanglement, i.e.,  von Neumann entropy of either of the
subsystems. For mixed states, it is given by
\be
E_f(\varrho)=\min \sum_ip_iE_f(\psi_i),\ \ \mbox{with} \  \
\varrho=\sum_ip_i|\psi_i\ra\la\psi_i|
\ee
where the minimum is taken over all possible decompositions of $\varrho$
(we call the decomposition realizing the minimum the optimal decomposition of
$\varrho$). Now $E_F\equiv \lim_n E_f(\varrho^{\otimes n})/n$.

To define distillable entanglement $E_D$ \cite{huge,Rains_bound1}
(see Ref. \cite{Rains_rig} for justifying
this definition) of the state $\varrho$ we consider distillation
protocols $\pcal$ given by a sequence
of trace preserving, completely positive  superoperators $\Lambda_n$,
that can be realized by using LQCC operations, and that map the state
$\varrho^{\otimes n}$ of $n$ input pairs  into a state $\sigma_n$ acting
on the Hilbert space $\hcal^{out}_n=\hcal_n\otimes \hcal_n$ with
$\dim \hcal_n=d_n$.
Define the maximally entangled state on the space $\hcal\otimes \hcal$ by
\be
P_+(\hcal)=|\psi_+(\hcal)\ra\la\psi_+(\hcal)|,\quad
\psi_+(\hcal)={1\over \sqrt{d}}\sum_{i=1}^{d} |ii\ra
\ee
where $|i\ra$ are basis vectors in $\hcal$, while $d=\dim\hcal$.
Now $\pcal$ is distillation protocol if, for high $n$, the final state
approaches the above  state $P_+$,
\be
F\equiv \la\psi_+(\hcal_n)|\sigma_n| \psi_+(\hcal_n)\ra \rightarrow 1
\ee
(i.e. the fidelity $F$ tends to 1). The asymptotic ratio $D_\pcal$ of
distillation via protocol $\pcal$ is given by
\be
D_{\pcal}(\varrho)  \equiv \lim_{n\rightarrow \infty}
 {\log_2\dim \hcal_n \over n}
\ee
The distillable entanglement is defined by thethe  maximum of $D_{\pcal}$ over
all protocols
\be
E_D(\varrho)=\sup_{\pcal} D_{\pcal}.
\ee

Now, the main result of this paper is the following.

{\bf Theorem 1.} For any function $E$ satisfying the introduced postulates,
and for any state $\varrho$,
one has
\be
E_D(\varrho)\leq E(\varrho)\leq E_F(\varrho).
\ee

{\bf Remark.} For pure states we have $E_D=E_F$; hence from the above
inequality it follows that all measures are equal to $E_F$ in this case. This
is compatible with the uniqueness theorem.

{\bf Proof.} Surprisingly enough,  the proof is elementary. Both left- 
 and right-hand-side inequality of the theorem are proved by the use of
the same line of argumentation:
\bei
\item by definition $E_D$ ($E_F$) is
asymptotically constant  during optimal distillation (formation) protocol
\item distillation (formation) protocol is an LQCC operation and cannot
increase any entanglement measure
\item the final (the initial) state is the pure one
\item for pure states all measures coincide by virtue of uniqueness theorem
\eei
Then it easily follows that, if  the given measure $E$ were, e.g,. less than
$E_D$, it would have to increase under optimal distillation protocol.
We used here
additivity, because formation and distillation protocols are collective
operations (performed on $\varrho^{\otimes n}$). Continuity is needed,
because we use the uniqueness theorem. By
writing the above more formally  in the case  $E\leq E_F$ we obtain:
\ben
&&E(\varrho)={E(\varrho^{\otimes n})\over n} \leq{\sum_ip_i E(\psi_i)\over n}
={\sum_ip_i E_f(\psi_i)\over n}\nonumber \\
&&={E_f(\varrho^{\otimes n})\over n}\mathop{\rightarrow}\limits^{n\rightarrow
\infty}
E_F(\varrho)
\label{dowod_ef}
\een
where we chose optimal decomposition of $\varrho^{\otimes n}$, so that
the $\sum_ip_i E_f(\psi_i)$ is minimal and hence equal to $E_f(\varrho^{\otimes
n})$ \cite{roofs}.
The  first equality comes from additivity, the inequality is a consequence of
monotonicity (more precisely - convexity,  axiom 2b)). The next-to-last
equality
follows from the   uniqueness theorem.
We will  skip the formal proof of the inequality $E_D\leq E$, because in the
following we prove formally a stronger result concerning
bounds for entanglement of distillation.  \blacksquare

Below we will show that the above, very transparent line of argumentation
is a powerful tool, as it allows to prove a very general theorem on
upper bounds of $E_D$.

{\bf Theorem 2.} Any function $B$ satisfying the
conditions a)-c) below is an upper bound for entanglement of distillation:

{\parindent=0pt
a) Weak monotonicity: $B(\varrho)\geq B(\Lambda(\varrho))$ where
$\Lambda$ is  the trace-preserving superoperator realizable
by means  of LQCC operations.
\vskip1mm
b) Partial  subadditivity: $B(\varrho^{\otimes n})\leq nB(\varrho)$
\vskip1mm
c) Continuity for isotropic state $\varrho(F,d)$ \cite{xor,Rains_bound2}.
The latter is of the form
\be
\varrho(F,d)= p P_+(C^d) + (1-p) {1\over d^2}I,\quad 0\leq p\leq1
\ee
with $\trace \big[\varrho(F,d) P_+(C^d)\big]=F$.
Suppose now that we have a sequence of isotropic states $\varrho(F_d,d)$,
such that $F_d\rightarrow 1$ if $d\rightarrow \infty$.
Then we require
\be
\lim_{d\rightarrow \infty} {1\over \log_2 d} B(\varrho(F_d,d))\rightarrow 1.
\ee}

{\bf Remarks.} (1) The above conditions are implied by
our postulates for entanglement measures. Specifically:
the condition a) is implied by monotonicity, b),  by additivity,
while the condition c), by continuity plus additivity. (2)
If instead of LQCC operations we take other class $C$ of operations
including
one-way classical communication, the {\it mutatis mutandis} proof
also applies (then the condition a) would involve the class $C$).


{\bf Proof.} We will perform analogous evaluation
as in formula (\ref{dowod_ef}) (now, however, we will not even use
the uniqueness theorem).
By subadditivity we have
\be
B(\varrho)\geq {1\over n}B(\varrho^{\otimes n}).
\label{dowod_ed1}
\ee
Since the only relevant parameters of the output of the process of distillation
are the dimension of the output Hilbert space and fidelity $F$ (see
definition of distillable entanglement), we can
consider distillation protocol ended by twirling \cite{xor}, that results in
isotropic final state. By condition a), distillation does not increase
$B$, hence
\be
{1\over n} B(\varrho^{\otimes n})\geq {1\over n} B(\varrho(F_{d_n},d_n))
\ee
Now, in the limit of large $n$, distillation protocol produces
$F\rightarrow 1$ and $(\log_2 d_n)/n \rightarrow E_D(\varrho)$; hence
by condition c) the right hand side of  the inequality tends to $E_D(\varrho)$.
Thus we obtain that $B(\varrho)\geq E_D(\varrho)$.\blacksquare

Using the above theorem, to find a bound for $E_D$, three things must be
done: one should show that a chosen function satisfies the weak monotonicity,
then check subadditivity and calculate it for isotropic state, to check
the condition c). Note that the weak monotonicity is indeed much easier
to prove than full monotonicity, as given by postulate 2a).
Checking subadditivity, in contrast to additivity,  is in many cases
immediate: It in fact holds for all so-far-known entanglement monotones.
Finally, the isotropic state is probably the easiest possible state to
calculate the value of a given function. To illustrate the power of the result
let us prove that relative entropy entanglement $E_r$ is bound for $E_D$.
Subadditivity, and weak monotonicity  are immediate consequence of the
properties of relative entropy used in definition of $E_r$
(subadditivity proved in Ref. \cite{Knight}, weak monotonicity --
in Ref. \cite{VP98}).
The calculation of $E_r$ for isotropic state is a little bit more involved,
but by using high symmetry of the state  it was found to be
\cite{Rains_bound2} $E_r(\varrho(F,d))= \log_2 d +F\log_2 F +
(1-F)\log_2{1-F\over d-1}$.
By evaluating  this expression now for large $d$, we easily obtain
that the condition c) is satisfied. The proof applies without any
change to the Rains  bound \cite{Rains_bound2}.


To summarize, we have presented two results. The first one has conceptual meaning
leading to deeper understanding the phenomenon of entanglement. It provides
some synthetic overview of the domain of quantifying entanglement in
asymptotic regime. One
of possible applications of the result would be to reverse the
direction of reasoning, and accept the condition $E_D\leq E\leq E_F$ as a
preliminary test for a good candidate for entanglement measure.
The second result presented in this paper is of direct practical use.
We believe that it will  make the search for strong bounds on $E_D$ much
easier, especially in higher dimensions. Finally we would like to stress
that the results display the power of the fundamental principle of
entanglement processing: the latter allow not only to replace a
complicated proof by a straightforward one, but also makes the
argumentation very transparent from the physical point of view.

We are grateful to E. Rains for stimulating discussion.
We would also like to thank the participants of the ESF-Newton
workshop (Cambridge 1999),
especially C. H. Bennett, S. Lloyd and G. Vidal for helpful
comments. The work is supported
by Polish Committee for Scientific Research, contract No. 2 P03B 103 16.


\begin{references}
\bibitem{roofs}
Note that this is a slight extension of the Uhlmann
argument concerning convex   roofs \cite{Uhlmann}.
\end{references}
\end{document}